\documentclass[11pt,a4paper]{article}
\usepackage{jcappub}

\usepackage{color}
\input{colordvi.tex}

\usepackage{amsmath}
\usepackage{graphicx}
\usepackage{float}
\usepackage{wrapfig}
\usepackage{bm}
\usepackage{amsmath}

\usepackage{amsmath,amssymb}
\usepackage{graphicx}
\usepackage{subfigure}
\usepackage{verbatim}
\usepackage{makeidx}

\usepackage{amssymb}

\title{CMB $\mu$ distortion from primordial gravitational waves}

\author[a]{Atsuhisa~Ota,}
\author[b]{Tomo~Takahashi,}
\author[c,d]{Hiroyuki~Tashiro}
\author[a]{and Masahide~Yamaguchi}
\affiliation[a]{Department of Physics, Tokyo
Institute of Technology,\\ Tokyo 152-8551, Japan}
\affiliation[b]{Department of Physics, Saga
University, \\Saga 840-8502, Japan}
\affiliation[c]{Department of Physics and
Astrophysics, Nagoya University,\\
 Nagoya 464-8602, Japan}
\affiliation[d]{Program for Leading Graduate Schools
\\``PhD
Professional:Gateway to Success in Frontier Asia'',}
\emailAdd{a.ota``at"th.phys.titech.ac.jp}
\emailAdd{tomot``at"cc.saga-u.ac.jp}
\emailAdd{hiroyuki.tashiro``at"nagoya-u.jp}
\emailAdd{gucci``at"phys.titech.ac.jp}
\abstract{
We propose a new mechanism of generating the $\mu$ distortion in cosmic
microwave background (CMB) originated from primordial gravitational
waves. Such $\mu$ distortion is generated by the damping of the
temperature anisotropies through the Thomson scattering, even on scales
larger than that of Silk damping. This mechanism is in  sharp
contrast with that from the primordial curvature (scalar) perturbations,
in which the temperature anisotropies mainly decay by Silk
damping effects. We estimate the size of the $\mu$ distortion from the
new mechanism, which can be used to constrain the amplitude of
primordial gravitational waves on smaller scales independently from the
CMB anisotropies, giving more wide-range constraint on their spectral
index by combining the amplitude from the CMB anisotropies.
}
\keywords{CMB $\mu$ distortion, gravitational wave}
\arxivnumber{1406.0451}

\begin{document}
\maketitle

\section{Introduction}

Cosmic microwave background (CMB) is one of the most useful remnants to
probe the early Universe. The recent observations of the CMB
anisotropies such as WMAP \cite{Hinshaw:2012aka} and Planck
\cite{Ade:2013uln} satellites strongly support the presence of the
accelerated period in the early Universe called inflation
\cite{inflation} and confirm that the primordial curvature perturbations
are almost scale-invariant, adiabatic, and Gaussian on large scales. In
addition, very recently, the BICEP2 collaboration \cite{Ade:2014xna} reported the
existence of the primordial gravitational waves (tensor perturbations),
which can determine the energy scale of inflation directly.

The spectral distortion in the CMB spectrum is another powerful tool to
probe phenomena in the early Universe. There are typically two types
of distortions, $\mu$- and
$y$-types~\cite{Zeldovich:1969ff,1970Ap&SS...7...20S}.  The $\mu$-type
distortion is a thermal distortion from the Planck distribution
characterized by non-zero chemical potential.  This kind of distortion
is mainly formed at the Compton equilibrium era, $5\times 10^4<z<10^7$ with $z$
being the redshift, because photon number conservation and non-zero
energy transfer under thermalization processes are indispensable for its
generation
\cite{Danese:1982,1991A&A...246...49B,Hu:1994bz,Chluba:2006kg,Khatri:2012tv}. Thus, one can
probe energy injection processes during this epoch through the non-zero
chemical potential $\mu$.  On the other hand, the $y$-type distortion is
a non-thermal type distortion relevant to the late epoch, $z<5\times 10^4$, in
which the Compton scattering is no longer effective to establish the
thermal equilibrium.  However, heated electrons can up scatter the CMB
photons via Compton scattering with energy transfer, which makes the CMB
spectrum deviate from the blackbody. This type of distortions is called
$y$-type and can probe energy injection processes at the late eras.  The
current constraints on $\mu$- and $y$-type distortions are obtained by
COBE FIRAS as $\mu< 9 \times 10^{-5}$ (95\% CL) and $y < 1.5 \times
10^{-5}$ (95\% CL)~\cite{Mather:1993ij,Fixsen:1996nj}, respectively.
Recently, future space missions such as PIXIE~\cite{Kogut:2011xw} and
PRISM~\cite{Andre:2013afa} are proposed and they have the potential to
detect the CMB distortions with $\mu \sim 10^{-8}$ and $y \sim
10^{-9}$. Therefore, it is expected that the constraint on the CMB
distortions will be significantly improved in the future.  In this
paper, we mainly concentrate on the first one, that is, the $\mu$-type
distortion.

One of the main mechanisms to generate the CMB $\mu$ distortion is via
Silk damping of the CMB acoustic waves~\cite{Hu:1994bz,Sunyaev:1970eu,
1991MNRAS.248...52B,1991ApJ...371...14D}.  During the tight
coupling epoch, the photon-baryon plasma can be regarded as a single
component and begins to oscillate together after the horizon entry.
However, once the coupling becomes weak, the ideal fluid approximation
gets worse and anisotropic stress becomes manifest, which generates
viscosity and causes the dissipation of the acoustic waves by Silk
damping~\cite{Silk:1967kq}.  Then, subsequent thermalization processes
realize new thermodynamic distribution of the CMB, which causes the
$\mu$ distortion. In a microscopic view, this is the mixing of the
different temperatures due to the diffusion of photons coming from
different phases of the acoustic waves, because the acoustic waves
induce the temperature fluctuations~\cite{Chluba:2004cn,Khatri:2012rt}. It is now
estimated that $\mu$ distortions are positive and the order of $10^{-8}$
for the curvature perturbations with almost scale invariant power
spectrum ${\cal P}_{\mathcal{R}}\sim 2.4\times 10^{-9}$
\cite{Hu:1994bz,Chluba:2012gq}.  In the case of
isocurvature perturbations, the distortion can be $10^{-11}$ for
neutrino isocurvature perturbations \cite{Chluba:2013dna}, and
$10^{-17}$ for CDM isocurvature perturbations
\cite{Chluba:2013dna,Dent:2012ne} with maximum amplitude allowed from
current constraint and almost scale invariant spectral index.  Thus, the
spectral distortions of the CMB are powerful tools to probe the
primordial perturbations, especially sensitive to those at smaller
scales.

In this paper, we propose a new generation mechanism of the CMB
distortion coming from the primordial tensor perturbations. The
primordial tensor perturbations generate the CMB temperature
fluctuations once they enter the horizon. However, these fluctuations
are damped through the Thomson scattering before the last scattering
surface, even on scales larger than Silk damping scale. This is in
sharp contrast with the CMB temperature fluctuations coming from the
curvature perturbations, which damps mainly below Silk damping
scale.  Although both processes 
correspond to the mixing of the blackbody spectra with the different
temperatures in the microscopic view, the mixing is due to the isotropic
nature of the Thomson scattering in the case with tensor perturbations unlike the diffusion of photons by Silk
damping as mentioned above. Thus, the CMB distortions in this
mechanism can be created even on scales larger than Silk damping
scale\footnote{Although this generation mechanism can create the CMB
distortion originated from the curvature perturbations, the generated
distortions can be dominated by the distortions due to Silk
damping}.
 The resultant spectrum after the mixing suffers from the thermalization
processes and ends up with the blackbody, Bose distribution with $\mu$ distortion or 
non thermal spectrum with $y$ distortion,  depending on the epoch of the mixing\footnote{See
Refs.~{\cite{Chluba:2004cn,Khatri:2012rt,Chluba:2012gq}} for generic discussions on the
generation of the CMB distortions from mixing of blackbodies.}. Then, we will estimate how much the $\mu$ distortions are
generated through such processes, which can be used to constrain the
amplitude of primordial gravitational waves (tensor perturbations) on
small scales, independently of other constraints. By combining the
amplitude probed by the CMB anisotropy experiments, it gives more
wide-range constraint on the spectral index of primordial tensor
perturbations using information on smaller scales.

The organization of this paper is as follows.  After briefly reviewing
the basics of the CMB $\mu$ distortion in the next section, we derive
the evolution equation of the CMB $\mu$ distortion coming from the
tensor perturbations in the section 3. Section 4 is devoted to the
concrete evaluation of the size of such $\mu$ distortion, given a
tensor-to-scalar ratio $r$ with a (constant) spectral index. We give the
conclusion and discussions in the final section.

\section{Basics of CMB $\mu$ distortion}

Even though the intensity of photon (temperature) spatially fluctuates,
we assume that the system is locally in thermal equilibrium. Then, the
distribution function at some space-time point ${\bm x}$ can be
parametrized as
\begin{align}
f({\bm x},\omega)=\frac{1}{e^{\frac{\omega }{T_\text{BE}({\bm x})}+\mu({\bm x})}-1}\label{36},
\end{align}
where $T_\text{BE}$ and $\omega$ are the local temperature and the
frequency of photons, respectively. The energy and the number densities
of photons with non-zero chemical potential, $\mu({\bm x})$, are given
by
\begin{align}
\rho({\bm x})&=\alpha T_{\text{BE}}^4({\bm x})\left(1-\frac{90\zeta(3)}{\pi^4}\mu({\bm x})\right),\\
n({\bm x})&=\beta T^3_{\text{BE}}({\bm x})\left(1-\frac{\pi^2}{6\zeta(3)}\mu({\bm x})\right),
\end{align}
respectively, where $\alpha$ and $\beta$ are some numerical constants.

We define the ``reference temperature'' and the ``reference Planck
distribution'' in terms of the second-order temperature perturbations by
equating the entropies of the photon fluid, that is, the number
densities of photons for both the thermal Bose-Einstein distribution and
the reference Planck one (for the details, see
Appendix~\ref{appendix})\footnote{
Even if we define it by equating the energy density, the final result
remains unchanged.
}. The reference temperature $T_\text{rf}$ is defined as
$(\beta^{-1}\langle n\rangle)^{1/3}$, where $\langle n\rangle$ denotes
the ensemble averaged number density. Therefore, $T_\text{rf}$ always
satisfies $T_\text{rf} \propto a^{-1}$ in the adiabatic expansion
case. Accordingly, the thermodynamical identity,
\begin{align}
\langle s\rangle =\frac{\langle \rho\rangle +\langle \mathcal{P}\rangle }{T_\text{rf}}
\end{align}
imposes that $\langle \rho\rangle \propto T_\text{rf}^4$ as well.  On
the other hand, the local Bose-Einstein temperature $T_\text{BE}$ is
divided into four parts:
\begin{align}
T_\text{BE}({\bm x})&=T_\text{pl}({\bm x})+t_\text{BE}({\bm x})\notag \\
&=\langle T_\text{pl}\rangle +\delta T({\bm x})+t_\text{BE}({\bm x})\notag \\
&=T_\text{rf} +\Delta T +\delta T({\bm x})+t_\text{BE}({\bm x}), 
\label{eq:local_bose_temp}
\end{align}
where $t_\text{BE}$ denotes the difference between the local
Bose-Einstein temperature and the local Planck temperature, and $\delta
T({\bm x})$ is the inhomogeneous part of local Planck
temperature. $\Delta T$ represents the difference between the averaged
Planck temperature $\langle T_\text{pl}\rangle$ and the reference
temperature $T_\text{rf}$, which is a second-order quantity of the
temperature fluctuation, as shown in Appendix \ref{appendix}. Due to
this difference, the averaged Planck temperature does not evolve as
$\langle T_\text{pl} \rangle \propto a^{-1}$ at the second-order
perturbation.

To simplify the following calculations, let us take dimensionless
temperature perturbations as $T_\text{rf} +\delta T({\bm x})+\Delta T
+t_\text{BE}({\bm x})=T_\text{rf}(1+\Theta({\bm x})+\Delta +t({\bm x}))$.  In this
notation, the number
 and energy densities can be rewritten as
\begin{align}
n({\bm x})&=\beta T_\text{rf}^3\left(1+3\Theta({\bm x}) +3\Delta +3t({\bm x}) +3\Theta^2({\bm x})-\frac{\pi^2}{6\zeta(3)}\mu({\bm x})\right),\\
\rho({\bm x})&=\alpha T_\text{rf}^4\left(1+4\Theta({\bm x}) +4\Delta +4t({\bm x}) +6\Theta^2({\bm x})-\frac{90\zeta(3)}{\pi^4}\mu({\bm x})\right),\label{2.7}
\end{align}
up to the second order of temperature perturbations. Note that only
$\Theta(\bm x)$ is the first order quantity in the above equations. By
imposing the number conservation with the adiabatic expansion,
$a^3\langle n \rangle$ should be constant, which leads to the following
equation:
\begin{align}
\Delta +\langle t\rangle =\frac{\pi^2}{18\zeta(3)}\langle \mu\rangle -\langle \Theta^2\rangle.
\end{align}
Here we have used that $T_\text{rf}$ scales as $a^{-1}$ and expanded up
to the second order in terms of $\langle \Theta^2\rangle.$ Substituting
the above equation to Eq.~(\ref{2.7}) yields the ensemble average of the
energy density as
\begin{align}
\langle \rho \rangle = \alpha T_\text{rf}^4\left[1+2\langle \Theta^2\rangle +\left(\frac{2\pi^2}{9\zeta(3)}-\frac{90\zeta(3)}{\pi^4}\right)\langle \mu\rangle \right].\label{2.10}
\end{align}
Since $\langle\rho\rangle \propto a^{-4}$, multiplying both sides of
Eq.~(\ref{2.10}) by $a^{-4}$ and taking the conformal time derivative,
we obtain the following formula for the evolution of the average
chemical potential $\langle \mu \rangle$,
\begin{align}
\label{eq:diff_mu}
\frac{d}{d\eta} \langle \mu\rangle =-1.4 \times 4\langle \Theta
 \dot \Theta\rangle +\mathcal O(\Theta^3),
\end{align}
where the dot represents the ordinary derivative with respect to the
conformal time $\eta$ and the ensemble average $\langle \mu \rangle$ can
be replaced by the spatial average.  By taking into account the
relaxation of the chemical potential due to the double Compton
scattering, we add a new term to the above formula:
\begin{align}
\frac{d}{d\eta} \langle \mu \rangle =-\frac{\mu}{t_\mu}-1.4 \times
 4\langle \Theta \dot \Theta\rangle +\mathcal O(\Theta^3).
\label{eq:diff_mu2}
\end{align}
Here $t_\mu$ is the decreasing time scale of $\mu$ due to the double
Compton scattering, which is given by~\cite{Hu:1992dc}
\begin{align}
t_\mu&=2.06\times 10^{33}(\Omega_bh^2)^{-1}\left(1-\frac{Y_\text{p}}{2}\right)^{-1}(1+z)^{-\frac{9}{2}}[\text{sec}].
\end{align}
The solution of Eq.~\eqref{eq:diff_mu2} can be formally expressed as
\begin{align}
\langle \mu\rangle=-1.4 \times 4\int^{\eta_{\rm fr}} _{0}d\eta'\mathcal J_{DC}(\eta')\langle \Theta \dot \Theta  \rangle,\quad \mathcal J_{DC}(\eta')=\exp{\left(-\int ^{z(\eta')}_{z(\eta=0)}\frac{dz}{{z}^3t_\mu(z)}\right)}\label{keisikikai},
\end{align}
where $\eta_{\rm fr}$ is the freeze-out epoch of a Bose-Einstein
distribution due to the Compton scattering~\cite{1991A&A...246...49B}.

\section{Boltzmann-Einstein system}

As shown in Eq.~(\ref{keisikikai}), the chemical potential depends on
the evolution of the temperature fluctuations. The CMB temperature
fluctuations can be created from the primordial perturbations generated
during  inflation. So, we briefly discuss the primordial
perturbations in this section.

The primordial perturbations produced during inflation can be classified
into two modes, the scalar one (primordial curvature perturbations) and
the tensor one (primordial gravitational waves). Since the CMB
distortions originated from the scalar mode are well investigated in the
context of Silk damping
\cite{Hu:1994bz,Chluba:2012gq,Chluba:2013dna,Dent:2012ne},
we focus on the tensor perturbations in this paper.

The tensor mode of the metric perturbations is given by the transverse
traceless component $H^{TT}_{ij}$ with
$\partial_iH^{TT}_{ij}=H^T_{ii}=0$ as
\begin{align}
ds^2&=-a^2d\eta^2+a^2\left(\delta_{ij}+H^{TT}_{ij}\right)dx^idx^j \label{eq:ichiyotoho}.
\end{align}
The Fourier component of $H^{TT}_{ij}$, which we denote by $\tilde
H^{TT}_{ij}$ in the following, can be decomposed as
\cite{Kosowsky:1994cy}
\begin{align}
\tilde H^{TT}_{ij}=h^+e^+_{ij}+h^\times e^\times_{ij},
\end{align}
where $e^A_{ij} ~(A=+,\times)$ are polarization bases for the plus and
the cross modes of the gravitational waves, respectively. Taking the
momentum of the gravitational waves parallel to the $z$ axis, the
polarization bases are given by
$e^+_{xx}=e^\times_{xy}=e^\times_{yx}=-e^+_{yy}=1$ with the zeroes
otherwise. By perturbing the Einstein equation, we have the evolution equations for the tensor
perturbations as
\begin{align}
\partial_\eta^2h^A+2\mathcal H\partial_\eta h^A +k^2 h^A=16\pi G a^2\pi^A,
\label{eq:wave-eq}
\end{align}
where $\pi^A$ is anisotropic stress of fluid. The primordial
gravitational waves are generated during inflation and their (initial)
amplitudes are characterized by the power spectrum as
\begin{align}
\mathcal P_{h^+}=\mathcal P_{h^\times}=\frac{4\pi G H^2}{\pi^2}
\label{def_power},
\end{align}
where $H$ is the Hubble parameter during inflation. The power spectrum
of the total tensor perturbations, $\mathcal P_T$, is related to these
power spectra as $\mathcal P_T =4\mathcal P_{h^+}=4\mathcal
P_{h^\times}$. It is commonly parametrized as
\begin{align}
\mathcal P_T&=rA_{\mathcal R}\left(\frac{k}{k_0}\right)^{n_T},
\end{align}
where $A_{\mathcal R}$ is the amplitude of the curvature power spectrum
at the pivot scale $k_0$, $r$ is the tensor-to-scalar ratio, and $n_T$
is the spectral index of the tensor perturbations. In this paper, we
adopt $A_{\mathcal R} =2.42\times 10^{-9}$ and
$k_0=0.002~\text{Mpc}^{-1}$ \cite{Hinshaw:2012aka}.

\subsection{Boltzmann equation}

Let us consider $2\times 2$ photon density matrix in Fourier space to
take into account photon polarization,
\begin{align}
f_{ij}=f^{(0)}\delta_{ij}+f^{(1)}_{ij},
 \label{eq:photon_matrix}
\end{align}
where $f^{(0)}$ is the background Planck distribution\footnote{
Note that
we do not need to take into account the difference between the Plank and
the Bose-Einstein distributions because such difference is manifest only
beyond the linear perturbation theory.
} and $f^{(1)}_{ij}$ is the
perturbed part. For convenience, we define $\Psi^T$ and
$\Psi^T_P$ as
\begin{align}
\Psi^T &= \frac{f^{(1)}_{11}+f^{(1)}_{22}}{2f^{(0)}}, \\
 \Psi^T_P &= \frac{f^{(1)}_{11}-f^{(1)}_{22}}{2f^{(0)}},
\label{eq:def_Psi}
\end{align}
which represent the perturbations for intensity and polarization
originated from primordial tensor perturbations\footnote{Strictly
speaking, $\Psi^T_P$ represents only the $Q$ component in the Stokes
parameter in our frame and there should be another quantity
corresponding to the $U$ Stokes component.  However, both quantities
obey the same equation (\ref{bol4}). Hence, we omit the latter for
simplicity.}. Each helicity 2 component (plus and cross) of $\Psi^T$ and
$\Psi^T_P$ is given by
\begin{align}
\Psi^T
&=(1-\lambda^2)\left(\Psi^{T+}\cos2\phi+\Psi^{T\times}\sin2\phi\right),\\
\Psi_P^T
&=(1+\lambda^2)\left(\Psi^{T+}_P\cos2\phi+\Psi^{T\times}_P\sin2\phi\right),
\end{align}
respectively, where $\lambda=\cos \theta$ and we have used the following
relations,
\begin{align}
\hat n_i\hat n_j e^+_{ij}&=\sin^2\theta \cos2\phi,\\
\hat n_i\hat n_j e^\times_{ij}&=\sin^2\theta \sin2\phi,
\end{align} 
for a photon direction vector with $\hat n_i
=(\sin\theta\cos\phi,\sin\theta \sin\phi,\cos\theta)$, and Fourier momentum is set to $k_\mu=(-k,0,0,k)$.
Following the
notation for the scalar perturbations given in Ref.~\cite{Ma:1995ey}, we
define the corresponding quantities originated from the tensor modes as
\begin{align}
F_{\gamma}=\frac{\int q^2dq qf^{(0)}(q)\Psi^T}{\int q^2dq
 qf^{(0)}(q)},
\label{eq:def_fg}
\\ 
G_{\gamma}=\frac{\int q^2dq qf^{(0)}(q)\Psi^T_P}{\int q^2dq qf^{(0)}(q)}.
\label{eq:def_gg}
\end{align}
It should be also noticed that $F_\gamma=4\Theta$ in the linear order.

$F_{\gamma}$ and $G_{\gamma}$ can be also decomposed into each
helicity 2 mode as
\begin{align}
F_{\gamma}&=(1-\lambda^2)\left(F^{T+}_{\gamma}\cos2\phi+F^{T\times}_{\gamma}\sin2\phi\right),\\
G_{\gamma}&=(1+\lambda^2)\left(G^{T+}_{\gamma}\cos2\phi+G^{T\times}_{\gamma}\sin2\phi\right).
\end{align}
Here, $F^{T+}_{\gamma}, F^{T\times}_{\gamma}, G^{T+}_{\gamma}$, and
$G^{T\times}_{\gamma}$ are defined in the similar way as
Eqs. (\ref{eq:def_fg}) and (\ref{eq:def_gg}) from $\Psi^{T+},
\Psi^{T\times}, \Psi^{T+}_P$, and $\Psi^{T\times}_P$, respectively. In
addition, in order to explicitly investigate the dependence of the
amplitude of primordial tensor perturbations, $F^{T+}_{\gamma}
(G^{T+}_{\gamma})$ and $F^{T\times}_{\gamma} (G^{T\times}_{\gamma})$ are
now normalized for $h^{+}=h^{\times}=1$.  By using the helicity 2
components, we have the following relation,
\begin{align}
F_\gamma \dot F_\gamma = (1-\lambda^2)^2
 \left[F^{T+}_\gamma \dot F^{T+}_\gamma \cos^22\phi
      +F^{T\times}_\gamma \dot F^{T\times}_\gamma \sin^22\phi
\right]+\cdots.\label{32}
\end{align}
Here the dots represent the contributions which would vanish after the
integration of $\phi$ and are irrelevant to the final estimate of the
$\mu$ distortion. This relation yields
\begin{align}
16\langle\Theta\dot{\Theta}\rangle =\int d\ln k\frac{\mathcal
 P_T(k)}{4}\int \frac{d\lambda}{2} \frac{d\phi}{2\pi}
 (1-\lambda^2)^2 \left[F^{T+}_\gamma \dot F^{T+}_\gamma \cos^22\phi
      +F^{T\times}_\gamma \dot F^{T\times}_\gamma \sin^22\phi
\right],
\label{8}
\end{align}
where we have used ${\mathcal P_T(k)}/4 = P_{h+}(k) = P_{h\times}(k)$.

The Boltzmann equations for $F_{\gamma}^{TA}$ and $G_{\gamma}^{TA}$
($A=+,\times$) are given by \cite{Kosowsky:1994cy,Bond:1984fp}
\begin{align}
\dot F^{TA}_\gamma&=\partial_\eta F^{TA}_\gamma+ik\lambda F^{TA}_\gamma+2\partial_\eta  h^{A}=
-\dot \tau(F^{TA}_\gamma-\Lambda^A),\label{bol3}\\
\dot G^{TA}_\gamma&=\partial_\eta G^{TA}_\gamma+ik\lambda G^{TA}_\gamma=-\dot \tau(G^{TA}_\gamma+\Lambda^A),\label{bol4}
\end{align}
where
\begin{align}
\Lambda^A&=\frac3{70} F^{TA}_{\gamma4}+\frac17
F^{TA}_{\gamma2}+\frac1{10} F^{TA}_{\gamma0}-\frac3{70}
G^{TA}_{\gamma4}+\frac67 G^{TA}_{\gamma2}-\frac35 G^{TA}_{\gamma0},
\label{eq:Lambda}
\end{align}
and, $F^{TA}_{\gamma l}$ and $G^{TA}_{\gamma l}$ in the right-hand side
of \eqref{eq:Lambda} are the multipole components of $F^{TA}_{\gamma}$
and $G^{TA}_{\gamma}$ as defined below.  It is now manifest that
$F^{T+}_\gamma (G^{T+}_\gamma)$ and $F^{T\times}_\gamma
(G^{T\times}_\gamma)$ obey the same equation with the same initial
amplitudes. Then, we can safely set $F^{T+}_\gamma = F^{T\times}_\gamma
\equiv F^{T}_\gamma$ and $G^{T+}_\gamma = G^{T\times}_\gamma \equiv
G^{T}_\gamma$ with $\Lambda^{+}=\Lambda^{\times}\equiv\Lambda$, which
yields
\begin{align}
16\langle\Theta\dot{\Theta}\rangle =\int d\ln k\frac{\mathcal
 P_T(k)}{4}\int \frac{d\lambda}{2} 
 (1-\lambda^2)^2 F^{T}_\gamma \dot F^{T}_\gamma.
\label{eq:8}
\end{align}
From Eq. (\ref{bol3}), we have the following equation,
\begin{align}
\int \frac{d\lambda}{2}
 (1-\lambda^2)^2 F^{T}_\gamma \dot F^{T}_\gamma
= -\dot{\tau} \int \frac{d\lambda}{2} 
 (1-\lambda^2)^2 \left[
F^{T}_\gamma F^{T}_\gamma -F^{T}_\gamma \Lambda
\right].
\label{82}
\end{align}

\subsection{Evaluation of the chemical potential $\mu$}

It is convenient to expand Eq.~(\ref{eq:8}) by multipoles with order of
$l$ because the Boltzmann equation can be solved order by order of
$l$. For the tensor components, we expand $F^T_\gamma$ and $G^T_\gamma$
as \cite{Ma:1995ey}
\begin{align}
  F^T_{\gamma}&=\sum_{l=0}(-i)^l(2l+1)P_l(\lambda)F^T_{\gamma l}, \\
  G^T_{\gamma}&=\sum_{l=0}(-i)^l(2l+1)P_l(\lambda)G^T_{\gamma l},
\end{align}
where $P_l(x)$ is the Legendre polynomial of order $l$. Here it should
be noticed that we expand $F^T_{\gamma}$ in stead of
$(1-\lambda^2)F^T_{\gamma}$. Therefore, the $l=0$ component,
$F^T_{\gamma 0}$, includes not only the monopole component but also the
quadrupole one because the factor $(1-\lambda^2)$ contains $P_2$ as well
as $P_0$.

The recursion relation of the Legendre polynomials,
\begin{align}
\lambda P_l=\frac{(l+1)P_{l+1}+lP_{l-1}}{2l+1},
\end{align}
yields
\begin{align}
(1-\lambda^2)P_l=A_lP_{l+2}+B_lP_l+C_lP_{l-2}\label{lgekousiki},
\end{align}
where $A_l$, $B_l$, $C_l$ are given by
\begin{align}
A_l=-\frac{(l+1)(l+2)}{(2l+1)(2l+3)},\quad B_l=\frac{2(l^2+l-1)}{(2l-1)(2l+3)},\quad C_l=-\frac{l(l-1)}{(2l+1)(2l-1)}.
\end{align}
This relation recasts the first term in the right-hand side
of Eq.~(\ref{82}) into
\begin{align}
\int\frac{d\lambda}{2}(1-\lambda^2)^2F^{T}_{\gamma}F^{T}_{\gamma}=\sum_{l=0}^{\infty}(-1)^l\Bigg[ a_l F^{T}_{\gamma l}F^{T}_{\gamma l}+b_lF^{T}_{\gamma l}F^{T}_{\gamma l+2}+c_lF^{T}_{\gamma l}F^{T}_{\gamma l+4}\bigg],\label{42}
\end{align}
where the coupling coefficients are expressed as
\begin{align}
a_l&=\frac{2(2l+1)(3l^4+6l^3-11l^2-14l+12)}{(2l-3)(2l-1)(2l+3)(2l+5)}, \\
b_l&=\frac{8(l+1)(l+2)(l^2+3l-2)}{(2l-1)(2l+3)(2l+7)}, \\
c_l&=\frac{2(l+1)(l+2)(l+3)(l+4)}{(2l+3)(2l+5)(2l+7)}.
\end{align}
Similarly, the second term in the right-hand side of Eq.~(\ref{82}) can
be rewritten as
\begin{align}
\int\frac{d\lambda}{2} (1-\lambda^2)^2F^{T}_{\gamma}\Lambda=\Lambda\int\frac{d\lambda}{2}\left(\frac{8}{35}P_4-\frac{16}{21}P_2+\frac{8}{15}P_0\right)F^{T}_\gamma=\Lambda \left(\frac{8}{35}F^{T}_{\gamma 4}+\frac{16}{21}F^{T}_{\gamma 2}+\frac{8}{15}F^{T}_{\gamma 0}\right).\label{47}
\end{align}

Therefore, the source term of the chemical potential for the tensor
modes is given up to the second-order perturbation by
\begin{align}
&\int\frac{d\lambda}{2}\frac{d\phi}{2\pi}(1-\lambda^2)^2F^T_\gamma \dot
 F^T_{\gamma} \nonumber
\\
&=-\dot \tau \bigg[
\frac{12}{25}F^T_{\gamma0}F^T_{\gamma0}+\frac{8}{25}F^T_{\gamma0}G^T_{\gamma0}+\frac{48}{35}F^T_{\gamma0}F^T_{\gamma2}-\frac{16}{35}F^T_{\gamma0}G^T_{\gamma2}-\frac{24}{35}F^T_{\gamma1}F^T_{\gamma1}+\frac{16}{35}F^T_{\gamma 2}G^T_{\gamma0}+\cdots \bigg],\label{tensor}
\end{align}
where the dots represent the contributions coming from higher order multipoles, which we can ignore safely.
Finally, inserting Eqs.~(\ref{eq:8}) and (\ref{tensor}) into
Eq.~(\ref{keisikikai}) yields the concrete expression for the $\mu$
distortion generated from the tensor modes up to the second order as
\begin{align}
&\mu^T=1.4\cdot  \frac14\int^{\eta_{\rm fr}}_{0}d\eta'\mathcal J_{DC}(\eta')\int d\ln k\frac{\mathcal P_T(k)}{4}\notag \\
\times \dot \tau &\bigg[
\frac{12}{25}F^T_{\gamma0}F^T_{\gamma0}+\frac{8}{25}F^T_{\gamma0}G^T_{\gamma0}+\frac{48}{35}F^T_{\gamma0}F^T_{\gamma2}-\frac{16}{35}F^T_{\gamma0}G^T_{\gamma2}-\frac{24}{35}F^T_{\gamma1}F^T_{\gamma1}+\frac{16}{35}F^T_{\gamma 2}G^T_{\gamma0}+\cdots\bigg].
 \label{eq:mu_tensor}
\end{align}
Only $F^T_{\gamma 0}$ couples to gravitational wave $h^A$ in the
Boltzmann hierarchies and thus it is generated from gravitational waves
directly while $G^T_{\gamma 0}$ is produced by $F^T_{\gamma 0}$ and
higher multipole components. Hence $F^T_{\gamma l}$ and $G^T_{\gamma l}$
with $l>0$ are created by the free streaming of the CMB photons.

\section{CMB $\mu$-distortion from primordial gravitational waves}

\subsection{Numerical results}

\begin{figure}
\centering
\includegraphics[width=12cm]{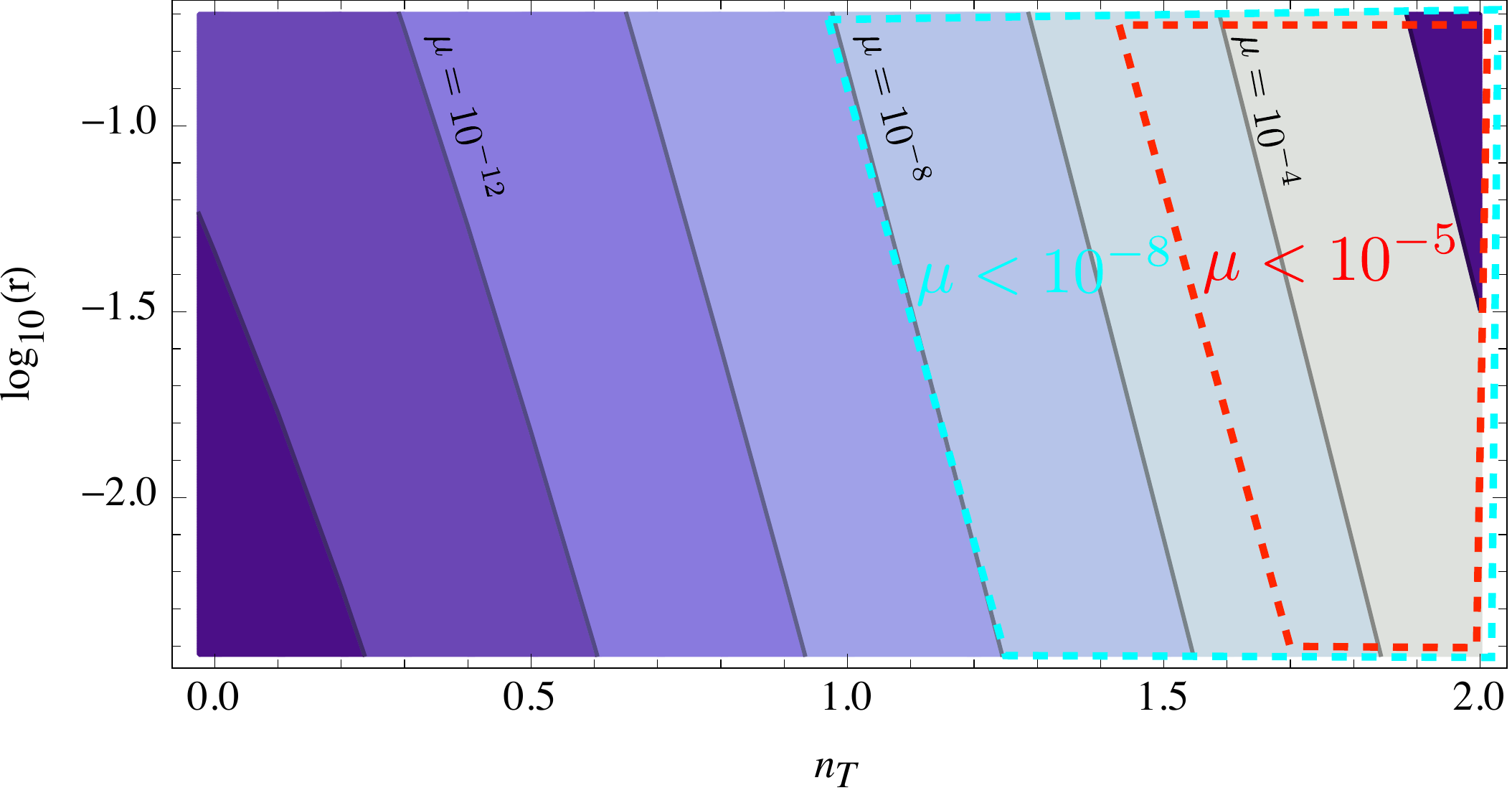}
\caption{The magnitude of the chemical potential $\mu$ generated from
primordial gravitational waves is shown in the $n_T$--$r$ plane.  The
regions enclosed by red dashed lines are ruled out by the COBE
satellite. Those by cyan dashed ones can be probed by the PIXIE.}
\label{nrtco}
\end{figure}

In this section, we numerically calculate Eq.~(\ref{eq:mu_tensor}) by
following the evolution of $F^T_{\gamma l}$ and $G^T_{\gamma l}$ using a
publicly available code,
CLASS~\cite{Lesgourgues:2011re,Blas:2011rf,Lesgourgues:2011rg,Lesgourgues:2011rh}.
The results are shown in Fig.~\ref{nrtco}, where contours of $\mu$ are
shown in the $n_T$--$r$ plane.  For $n_T=0$ and $r=0.2$, the generated
$\mu$ distortion is estimated as $\mu\sim 4.4\times 10^{-14}$. For
$n_T=1.0$ and $r=0.2$, the value of $\mu$ can be as large as $\mu\sim
1.4\times 10^{-8}$, which is comparable to that coming from the scalar
perturbations with $A_{\cal R} = 2.4 \times 10^{-9}$ and $n_s = 0.96$.
In fact, the BICEP2 data alone slightly prefers a blue-tilted
gravitational waves and such a blue spectral index is known to relax
the tension between the analyses from Planck temperature data and
BICEP2~\cite{Gerbino:2014eqa,Wang:2014kqa,Wu:2014qxa}.

The current constraint on the $\mu$ distortion given by COBE FIRAS is
$|\mu| < 9 \times 10^{-5}$ (95\% CL) \cite{Fixsen:1996nj}. This
constraint will be dramatically improved by future space mission such as
PIXIE or PRISM, e.g.~$|\mu| < 5 \times 10 ^{-8}$ by PIXIE at the $5
\sigma $ level \cite{Kogut:2011xw}.  For reference, the region ruled out
by the COBE satellite is enclosed by red dashed lines. The region probed
by PIXIE can be surrounded by cyan dashed lines.

Given the current constraint on $r \lesssim {\cal O}(0.1)$, primordial
gravitational waves with the scale invariant spectrum cannot produce
observable $\mu$ distortion. However, if their spectrum is significantly
blue-tilted, primordial gravitational waves can produce significant
$\mu$ distortion comparable to that from the scalar perturbations, which
implies that the future observations can provide the strong constraints
on $r$ and $n_T$.

Note that this consequence is the result of the indirect energy transfer
via CMB temperature fluctuations from gravitational waves. To understand
this, let us discuss Eq.~(\ref{bol3}) again. For simplicity, we neglect
the second terms, $ik\lambda F^T_\gamma$ and $\Lambda$, in both the
center and the right-hand side, which describe the free streaming of the
CMB photons and the anisotropic nature of Thomson scattering,
respectively. Since the Thomson scattering time scale is much shorter
than the cosmological time scale, we can take the steady state
approximation between the center and the right hand side and obtain
$2\partial_\eta h\sim- \dot \tau F^T_\gamma $. Accordingly we obtain
\begin{align}
\dot F^T_\gamma \sim- \dot \tau F^T_\gamma 
 \sim 2\partial_\eta h.\label{eq:42}
\end{align}
Eq.~(\ref{eq:42}) tells us that the temperature fluctuations created by
the integrated Sachs-Wolfe effect are decreasing during extremely short
time interval by the Thomson scattering. Therefore, since the generated CMB
distortion depends on how much CMB anisotropies are damped by the
Thomson scattering, the contribution to the distortions is proportional
to $F_{\gamma}\dot{F}_{\gamma}$.

\begin{figure}
\begin{center}
 \includegraphics[width=12cm]{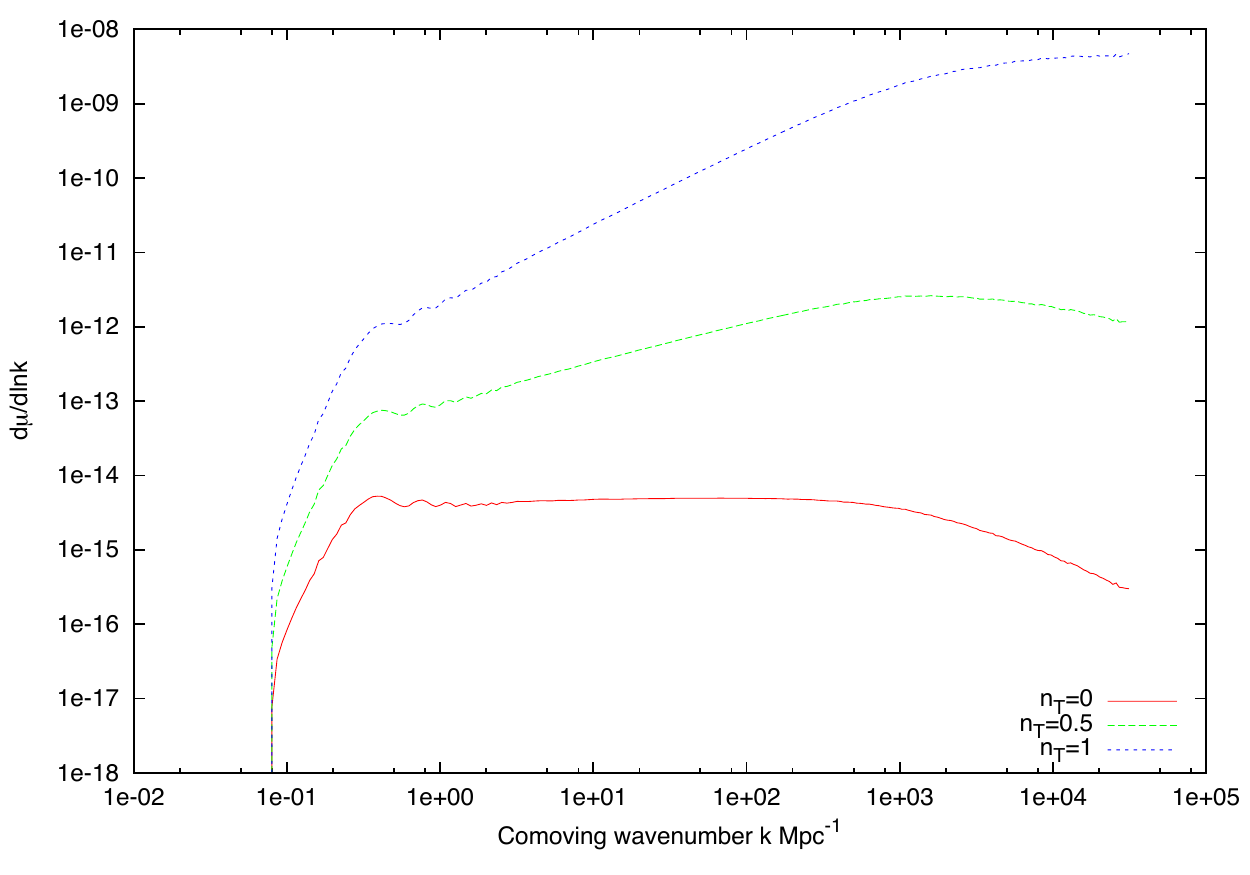}
\end{center}
\caption{The amount of $\mu$ generated per the logarithmic $k$ interval
originated from the tensor modes.  The red, green, and blue lines
correspond to the cases with $n_T =0$, $n_T=0.5$ and $n_T =1.0$,
respectively. \label{fig2}}
\end{figure}

\subsection{Comparison with the scalar perturbation case}

The CMB distortions originated from the scalar perturbations are mainly
generated by the energy release due to Silk damping.
In the similar way to Eq.~(\ref{eq:mu_tensor}), the chemical potential
created from the scalar perturbations can be calculated from
\cite{Khatri:2012rt,Chluba:2012gq}
\begin{align}
&\mu^S=1.4\cdot \frac14\int^{\eta_{\rm fr}}_{0}d\eta'\mathcal J_{DC}(\eta')
\int d(\ln k) \mathcal P_{\mathcal R}(k)\notag \\ \dot \tau &\bigg[
-\frac{4}{3k^2}(\Theta_e-\Theta_\gamma)^2-\sigma_\gamma(-18\sigma_{\gamma}+G^S_{\gamma2}+G^S_{\gamma0})+\sum_{l=3}
(-1)^{n}(2n+1)F^S_{\gamma l}F^S_{\gamma l}\bigg], \label{4.52}
\end{align}
where $F^S _{\gamma l}$ and $G^S_{\gamma l}$ are the Legendre-expansion
coefficients for the scalar  components of $F_\gamma$ and $G_\gamma$
defined in Eqs.~(\ref{eq:def_fg}) and (\ref{eq:def_gg}).
$\Theta_\gamma$, $\Theta_e$, and $\sigma_\gamma$ correspond to the
photon fluid velocity, the baryon fluid velocity and the anisotropic
stress of the photon fluid \cite{Ma:1995ey}, which are explicitly written as
\begin{align}
\Theta_\gamma  =\frac{3}{4} kF^S_{\gamma 1},\quad
\Theta_e  = ikv,\quad \sigma_\gamma=\frac{1}{2}F^S_{\gamma2},
\end{align}
with $v$ being the baryon fluid velocity potential.

For the case with the scalar mode, since the dominant generation
mechanism is Silk damping, the $\mu$ distortion is generated around
Silk damping scale~(see Fig.~4 in Ref.~\cite{Dent:2012ne}).  On
larger scales than Silk damping one, photon and baryon are tightly
coupled before the epoch of recombination. In the tight coupling
approximation, the velocity difference, $\Theta_e-\Theta_\gamma$ and the
anisotropic stress $\sigma_\gamma$ is of the order of $k/\dot \tau \ll
1$. Therefore, the contributions from such large scales are negligible.

On the other hand, since the temperature fluctuations due to the tensor
modes do not couple with the baryon fluids, they are not suppressed by
the order of $k/\dot \tau$. As mentioned above, the temperature
fluctuations can be approximated to $F^T_\gamma \sim -2\partial_\eta
h/\dot \tau$ over all scales. Therefore, the generation of the chemical
potential due to the tensor modes occurs even on larger scales, compared
with Silk damping case.

Fig.~\ref{fig2} shows that scale dependence of $\mu$ distortion. In the
figure, the vertical axis represents ${d\mu}/{d \ln k}$.  Compared to
the scalar mode cases~(e.g. Fig.~4 in Ref.~\cite{Dent:2012ne}), the
$\mu$ distortion generated from the tensor mode comes even from larger
scales. In particular, as one can see from Fig.~\ref{fig2}, the
contribution to the chemical potential dramatically increases around $k
= {\cal O}(0.1)$ Mpc$^{-1}$, which corresponds to the Horizon scale at
the epoch $\eta_{\rm fr}$. This is because, after horizon crossing, the
gravitational waves start to decay and produce the temperature
fluctuations through the integrated Sachs-Wolfe effect.
Fig.~\ref{fig2} also shows the cases for other values of  $n_T$.
As expected, the contribution from small scales increases for larger $n_T$.

\section{Conclusions and discussion}

In this paper, we have investigated CMB $\mu$ distortion originated from
primordial gravitational waves. The temperature anisotropies generated
from those are damped through the Thomson scattering, even on scales
larger than Silk damping scale, which leads to the generation of the
non-zero chemical potential $\mu$. Unfortunately, given the
tensor-to-scalar ratio of the order of unity and the scale invariance of
primordial gravitational waves, the created chemical potential $\mu$ is
as small as $10^{-13}$. However, once the blue spectral index for tensor
perturbations is allowed, the significant $\mu$ distortion can be
generated, which in turn strongly constrains the tensor-to-scalar ratio
and the tensor spectral index.

This new mechanism is quite different from that generated from the
scalar perturbations, in which Silk damping effects mainly damp the
temperature anisotropies while the damping is ineffective in the tight
coupling region. Thus, the chemical potential $\mu$ can be produced even
on larger scales for the tensor perturbations while that from the scalar
perturbation is created mainly below Silk damping scale. This kind
of the scale dependence may enable us to discriminate whether $\mu$
distortion is created from the tensor or the scalar perturbations, even
if the former is much smaller than the latter. This is because, if we
consider the cross correlation between the temperature anisotropies and
the (scale dependent) $\mu$ distortion, their correlated bispectra can
be large for the tensor mode compared to the scalar one since the tensor
mode is dominant on scales larger than Silk damping scale while that
from the scalar mode is significantly suppressed on such larger scales.
We will study this issue in the future work.

\acknowledgments 
We would like to thank Jens Chluba for his helpful comments.
This work is supported by the Japan Society for Promotion of Science
(JSPS) Grant-in-Aid for Scientific Research (Nos.~23740195 [TT],
25287057 [HT], 25287054 and~26610062 [MY]).


\appendix

\section{Mixing of Blackbodies}
\label{appendix}

\subsection{Mixing blackbodies to a new blackbody}

Here let us consider two blackbody spectra with different temperatures,
$T+\delta T$ and $T-\delta T$, respectively and mix them.  The total
energy density of this system is given by
\begin{align}
\rho_{\text{initial}}=\frac{\alpha}{2}[(T+\delta T)^4+(T-\delta T)^4]=\alpha  T^4\left[1+\frac32\left(\frac{\delta T}{T}\right)^2+\cdots\right]^4=\rho_{\text{final}},
\end{align}
which is conserved.
Assuming that the mixed system relaxes into one blackbody spectrum,
the final temperature of a new blackbody one is given by
\begin{align}
T_{\text{final}}= T\left[1+\frac32\left(\frac{\delta T}{T}\right)^2+\cdots\right].
\end{align}
In this case, we can easily confirm that the final number density
changes from the initial number density,
\begin{align}
n_{\text{initial}}=\beta T^3\left[1+3\left(\frac{\delta T}{T}\right)^2+\cdots\right]\to n_{\text{final}}=\beta T_{\text{final}}^3=\beta T^3\left[1+\frac{9}{2}\left(\frac{\delta T}{T}\right)^2+\cdots \right].
\end{align}
This result implies that, only when the number is not conserved, the new
system can relax into one blackbody.

\subsection{Mixing blackbodies under both of energy and number conservations}

In this subsection, we mix two blackbodies by imposing not only the
energy conservation but also the number conservation. In this setting,
as shown in the previous subsection, the new mixed system cannot relax
into a blackbody, instead, relax into the Bose-Einstein system with a
non-zero chemical potential.

For such a Bose-Einstein distribution, the energy and the number
densities with a non-zero chemical potential $\mu$ are estimated up to
the first order of $\mu$ as
\begin{align}
\rho&=\alpha T_{\text{BE}}^4\left(1-\frac{90\zeta(3)}{\pi^4}\mu\right)\\
n&=\beta T^3_{\text{BE}}\left(1-\frac{\pi^2}{6\zeta(3)}\mu\right),
\end{align}
where $T_\text{BE}$ is the temperature of this Bose-Einstein
distribution. From the number and the energy conservations, $\mu$ and
$T_\text{BE}$ are easily estimated as \cite{Khatri:2012rt},
\begin{align}
\mu&=\frac{1}{\left(\frac{\pi^2}{9\zeta(3)}-\frac{45\zeta(3)}{\pi^4}\right)}\left(\frac{\delta T}{T}\right)^2,\\
T_\text{BE}&=T\left[1+\frac{\frac{\pi^2}{6\zeta(3)}-\frac{45\zeta(3)}{\pi^4}}{\frac{\pi^2}{9\zeta(3)}-\frac{45\zeta(3)}{\pi^4}}\left(\frac{\delta T}{T}\right)^2\right].
\end{align}
It is now manifest that both of the temperature shift and the generated
chemical potential are of the second order in $\delta T/T$.

\end{document}